\documentclass[%
reprint,
superscriptaddress,
 amsmath,amssymb,
 aps,
]{revtex4-2}

\usepackage{graphicx}
\usepackage{dcolumn}
\usepackage{bm}

\begin{document}

\title{Aluminum nuclear demagnetization refrigerator for powerful continuous cooling}

\author{Matthias Raba}
\altaffiliation{Now at: DSBT Univ. Grenoble Alpes, CEA, CNRS, F-38000 Grenoble, France}
\affiliation{Univ. Grenoble Alpes, CNRS, Grenoble INP, Institut Néel, 38000 Grenoble, France}
\author{S\'ebastien Triqueneaux}
\affiliation{Univ. Grenoble Alpes, CNRS, Grenoble INP, Institut Néel, 38000 Grenoble, France}
\author{James Butterworth}
\affiliation{Air Liquide Advanced Technologies, 2, rue de Clemenciere, BP 15, 38360 Sassenage, France}
\author{David Schmoranzer}
\affiliation{Charles University, Ke Karlovu 3, 121 16 Prague, Czech Republic}
\author{Emilio Barria}
\affiliation{Univ. Grenoble Alpes, CNRS, Grenoble INP, Institut Néel, 38000 Grenoble, France}
\author{J\'er\^ome Debray}
\affiliation{Univ. Grenoble Alpes, CNRS, Grenoble INP, Institut Néel, 38000 Grenoble, France}
\author{Guillaume Donnier-Valentin}
\affiliation{Univ. Grenoble Alpes, CNRS, Grenoble INP, Institut Néel, 38000 Grenoble, France}
\author{Thibaut Gandit}
\affiliation{Univ. Grenoble Alpes, CNRS, Grenoble INP, Institut Néel, 38000 Grenoble, France}
\author{Anne Gerardin}
\affiliation{Univ. Grenoble Alpes, CNRS, Grenoble INP, Institut Néel, 38000 Grenoble, France}
\author{Johannes Goupy}
\affiliation{SPINTEC, Univ. Grenoble Alpes, CEA, CNRS, F-38000 Grenoble, France}
\author{Olivier Tissot}
\affiliation{Univ. Grenoble Alpes, CNRS, Grenoble INP, Institut Néel, 38000 Grenoble, France}
\author{Eddy Collin}
\affiliation{Univ. Grenoble Alpes, CNRS, Grenoble INP, Institut Néel, 38000 Grenoble, France}
\author{Andrew Fefferman}
\email{andrew.fefferman@neel.cnrs.fr}
\affiliation{Univ. Grenoble Alpes, CNRS, Grenoble INP, Institut Néel, 38000 Grenoble, France}
\date{\today}

\begin{abstract}
Many laboratories routinely cool samples to 10 mK, but relatively few can cool condensed matter below 1 mK. Easy access to the microkelvin range would propel fields such as quantum sensors and quantum materials. Such temperatures are achieved with adiabatic nuclear demagnetization. Existing nuclear demagnetization refrigerators (NDR) are “single-shot”, and the recycling time is incompatible with some sub-mK experiments. Furthermore, a high cooling power is required to overcome the excess heat load of order nW on NDR pre-cooled by cryogen-free dilution refrigerators. We report the performance of an aluminum NDR designed for powerful cooling when part of a dual stage continuous NDR (CNDR). Its thermal resistance is minimized to maximize the cycling rate of the CNDR and consequently its cooling power. At the same time, its susceptibility to eddy current heating is minimized. A CNDR based on two of the aluminum NDR presented here would achieve a cooling power of approximately 40 nW at 560 $\mu$K less than six days after cooling from room temperature, with a small offset in electronic temperature that decreases as the time-dependent heat load decays.
\end{abstract}

\maketitle


\section{\label{sec:intro} Introduction}
Cooling condensed matter well below the $\approx5$ mK limit of dilution refrigeration was motivated in part by the rich physics generated by the internal degrees of freedom of superfluid $^3$He. Studies of this system continue to generate fascinating new results \cite{Tian23,Makinen23,Autti21,Heikkinen21,Autti23b,Shook24}. Today, a broad spectrum of research is conducted near 1 mK and far below. These include quantum criticality in strongly correlated electron systems \cite{Schuberth16,Nguyen21}, nanomechanical resonators \cite{Cattiaux21}, amorphous solids \cite{Strehlow98,Fefferman08,Pedurand24}, electronic transport in nanostructures \cite{Sarsby20}, 2D electron gases \cite{Levitin22} and dark matter searches \cite{Chen23,Autti24}.

At present, microkelvin temperatures are limited to a small number of laboratories, including those of the European Microkelvin Platform \cite{Pickett18}. Such facilities employ adiabatic nuclear demagnetization refrigeration (NDR) to achieve these low temperatures. The refrigerant of a NDR is a metal, usually copper or PrNi$_5$, containing nuclei with non-zero spin. First, a magnetic field is applied to the refrigerant while it is thermally coupled to the pre-cooling dilution refrigerator at $\approx 10$ mK, causing the spins to align with the magnetic field. Then the refrigerant is thermally isolated from the precooler and the applied field is decreased. Since the thermal isolation implies that the entropy of the spins remains constant, the temperature of the spins ideally decreases in proportion to the field. Once the desired temperature is achieved, heat leaks are balanced by further demagnetization. In the best conventional systems the minimum heat leaks are $\approx100$ pW and are limited by energy deposition by cosmic rays \cite{Nazaretski04}.

Cryogen-free (``dry'') dilution refrigeration \cite{Prouve07} simplified access to temperatures near 10 mK and allowed broad development of superconducting quantum technology. Along with the lack of helium transfers, the dry environment allows for much more experimental space and easy automation. Developing similarly easy access to microkelvin temperatures is therefore a field of intense interest. This is achieved by pre-cooling NDR with dry dilution cryostats \cite{Batey13,Todoshchenko14,Palma17,Yan21,VanHeck23}. A state of the art cryogen-free system maintains temperatures below 1 mK for 140 h at a time under a 2 nW heat load \cite{Nyeki22}. While the sample must be warmed near 10 mK to recycle the system, it is possible to operate experiments below 1 mK with a 95\% duty cycle. Some researchers use on-chip nuclear demagnetization to cool electrically insulated samples, thereby overcoming the thermal bottleneck caused by weak electron-phonon coupling at the cost of large magnetic fields at the sample \cite{Samani22}. The performance of on-chip NDR would be enhanced by pre-cooling of the insulating chip by macroscopic NDR \cite{Autti23}.

Dry NDR systems experience relatively high heat leaks of at least 2 nW \cite{Nyeki22}. This is partly due to the vibrations of the pulse tube that replaces the liquid helium bath used in conventional refrigerators. The excess heat leak limits the time available for uninterrupted measurements at ultra-low temperatures and constrains the applicability of sub-mK refrigerators. In particular, systems with randomly configured degrees of freedom that may be reset by cycling between 10 mK and sub-mK temperatures cannot be studied if the experimental window is too short. One example is the search for individual intrinsic two level systems \cite{Ramos13}, where the energies of TLS resonant with a mechanical resonator may shift upon thermal cycling. Furthermore, systems that are weakly coupled to the cryostat experience thermal decoupling due to time dependent internal heat release. In particular, extremely high quality mechanical modes are, by definition, weakly coupled to their phonon baths. Therefore, a long cold-time may prove essential in achieving very long mechanical coherence times, which are at present limited by temperature dependent losses inside the mechanical elements \cite{Seis22}. Similar concerns may arise in future work on fluxonium qubits, which have transition frequencies in the low MHz range and therefore would benefit from cooling below 1 mK \cite{Nguyen19,Najera24}. Finally, the need to periodically recycle the cryostat may be considered too inconvenient to many workers, especially in the search for rare astrophysical events or in industrial settings \cite{Albash17}.

Continuous cooling below 1 mK using two NDR has been proposed \cite{Toda18,schmoranzer2019b}. The NDR stages may be arranged in a serial or parallel configuration \cite{Schmoranzer20}. One design for continuous NDR (CNDR) relies on PrNi$_5$ refrigerant, with a projected cooling power of 10 nW at 1 mK \cite{Takimoto22}. Advantages of this refrigerant include the enhancement of the magnetic field experienced by the nuclei to $\approx10$ times the applied one, the small Korringa constant $\kappa\lesssim 1$ mK sec (compared with 1.80, 1.27 and 1.09 K sec for $^{27}$Al, $^{63}$Cu and $^{65}$Cu, respectively) yielding rapid thermal equilibration of nuclei and electrons, and a high electrical resistance yielding low eddy current heating \cite{Andres82,Pobell07}. These features of PrNi$_5$ are balanced by its fragility \cite{Takimoto22}, difficulties in its synthesis that may lead to magnetic losses \cite{Andres82}, its high cost and limited availability, the minimum temperature $\approx0.4$ mK that is limited by magnetic ordering of the nuclei \cite{Kubota80}, and the low thermal conductance of the alloy. The thermal resistance of a PrNi$_5$ stage can be improved by soldering Cu or Ag conductors to the sides of the PrNi$_5$ rods. However, the 29 mT critical field of indium solder \cite{Nyeki22} and the similarly strong internal magnetic field of PrNi$_5$ \cite{Kubota80} limit the final applied field.

The cooling power of the CNDR at a given temperature increases with the thermal conductance between the nuclei of the two stages \cite{Schmoranzer20}. Butterworth \emph{et al.} demonstrated that the thermal resistance of a critical component of the thermal connection, the superconducting heat switch, can be reduced to a negligible value while maintaining a high switching ratio \cite{Butterworth22}. This was achieved by joining Cu to the Al superconducting element with minimal contact resistance. The capability to make low resistance Al/Cu joints makes aluminum attractive not only as a switching element but also as a nuclear refrigerant. This material is readily available in 6N purity, meaning that the thermal resistance of an Al stage can be made very low, thereby optimizing another part of the thermal link between the nuclei of the two stages. Furthermore, Al is abundant and inexpensive; it is weldable and flexible; its nuclei have spin 5/2; its magnetoresistance is acceptable; and nuclear magnetic ordering is expected to occur at a few $\mu$K \cite{Wendler97}, i.e., far below the 0.4 mK ordering temperature in PrNi$_5$. The literature discussing the use of aluminum for NDR is very limited. In \cite{Wendler97}, an Al stage was precooled by a Cu stage to 2.2 mK. Demagnetization of the Al stage resulted in a minimum electron temperature below 100 $\mu$K.

We report the performance of a new aluminum nuclear demagnetization refrigerator. The nuclear stage is composed of a bundle of 4N pure aluminum wires, which simultaneously optimizes the thermal conductance along the nuclear stage and the eddy current heating. The system operates at a relatively low maximum magnetic field of $\approx 3$ T. This allows efficient reduction of stray fields so that samples and thermometers sensitive to magnetic fields can be used. In particular, a SQUID-based magnetic field fluctuation thermometer (MFFT) \cite{MFFT,Engert12} was used in the present work. Furthermore, the size of the two main coils of the future CNDR will be relatively small and the interaction between the coils will be minimal.

\section{Design}
\begin{figure}
\includegraphics[width=\linewidth]{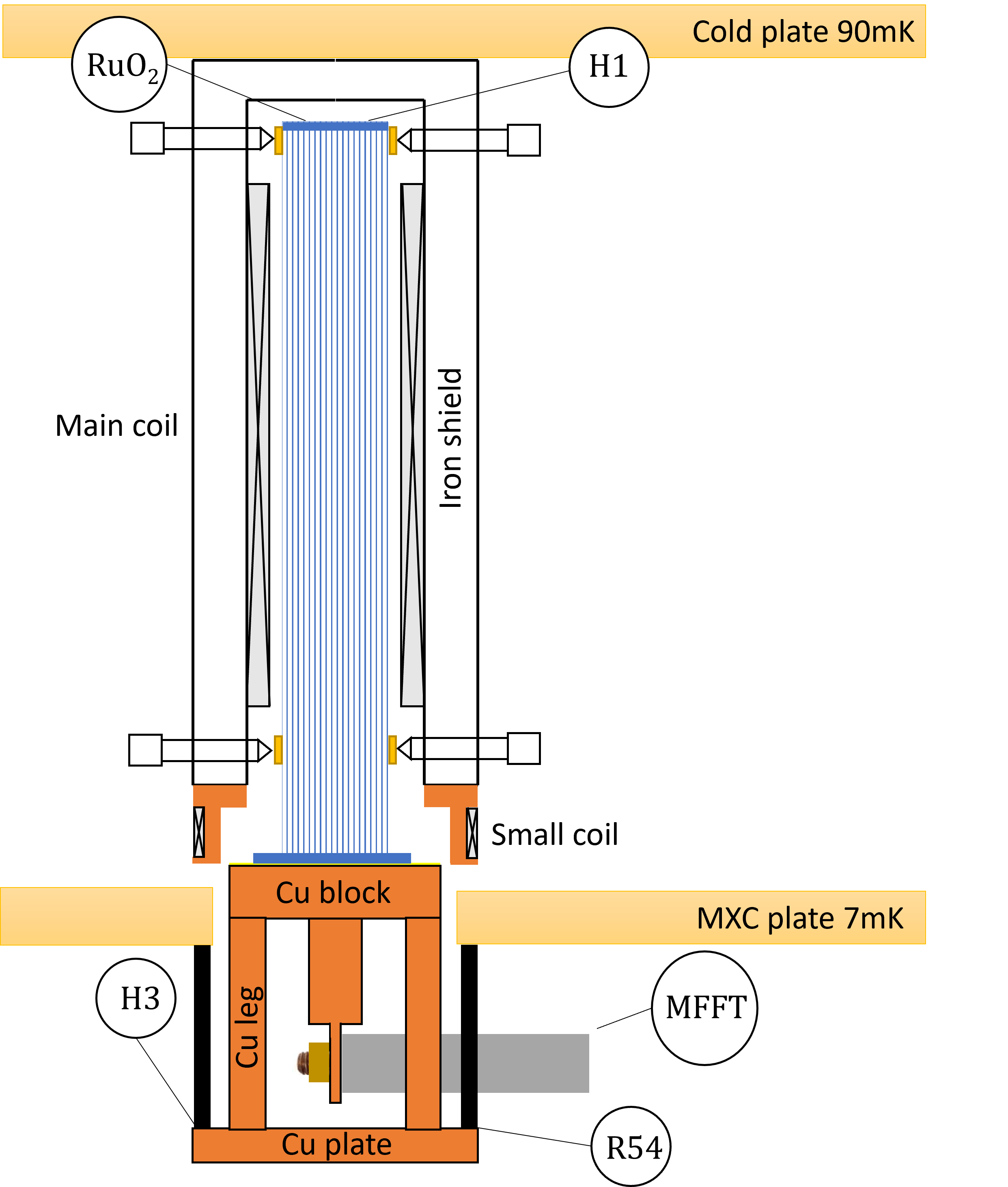}
\caption{\label{fig:design} Schematic of the aluminum wire demagnetization stage and magnets mounted on the dilution pre-cooler. The Cu plate is connected to the MXC plate via a heat switch (not shown). Metal film heaters H1 and H3 are placed at the top and bottom of the nuclear stage, a ruthenium oxide thermometer is at the top of the stage, and a carbon resistor R54 and magnetic field fluctuation thermometer are at the bottom of the stage. Nylon screws are used to center the Al wire bundle.}
\end{figure}

A schematic of the present system is shown in Fig. \ref{fig:design}. It is compatible with a standard cryogen-free dilution refrigerator, so that the temperature range of such a cryostat can be extended below 1 mK. In our case, a Bluefors LD-400 dilution refrigerator was used to precool the demagnetization stage. The windings of the main coil extended over 10 cm with a 22 mm bore diameter. Its maximum field was 2.84 T. The coil was surrounded by a 1 cm thick soft iron magnetic shield. The coil assembly was anchored to the bottom of an intermediate stage of the cryostat known as the cold plate. A small coil attached to the bottom of the main coil assembly was used to maintain the aluminum between the main coil and the copper thermal link in its normal state by constantly applying a magnetic field above the critical field $B_c=10$ mT. With the main coil fully energized, the magnetic field 4 cm below the mixing chamber plate was less than 2 mT.

The nuclear stage consisted of a bundle of 280 aluminum wires of 1 mm diameter and 4 N purity. It was centered inside the main coil with nylon screws. The wires were welded at the top and at the bottom of the bundle. The weld at the top provided a platform for a ruthenium oxide resistance thermometer and a heater H1. The weld at the bottom was gold plated and pressed to a gold plated copper block, using the technique described in \cite{Triqueneaux21}. The MFFT was attached to the copper block. The copper block was also connected to the heat switch described in \cite{Butterworth22} using a series of copper pieces, including a copper plate on which we mounted a carbon resistance thermometer R54 and a heater H3. The copper plate was attached to the mixing chamber plate with carbon rods. The other terminal of the heat switch was bolted to the top of the mixing chamber plate, whose temperature was monitored with another MFFT.

In the present work, we did not attempt to minimize the resistance between the copper block and the heat switch. That was not necessary to achieve our goal of demonstrating the suitability of the stage, consisting of the Al wire bundle and Au-plated Cu block, for a CNDR. We are confident that the thermal resistance of the thermal link outside of the main coil will not limit the cooling power of the future CNDR. In particular, this thermal link will consist of heat switches with a resistance of 3 n$\Omega$ \cite{Butterworth22}, Cu bars with a cross section of 1 cm$^2$ and a resistance per unit length of 0.3 n$\Omega$/cm, and Au-plated Cu/Cu joints. We have reproduced the $<10$ n$\Omega$ resistance of demountable Cu/Cu joints reported in \cite{Okamoto90}.

\section{Performance}

\begin{figure}
\includegraphics[width=\linewidth]{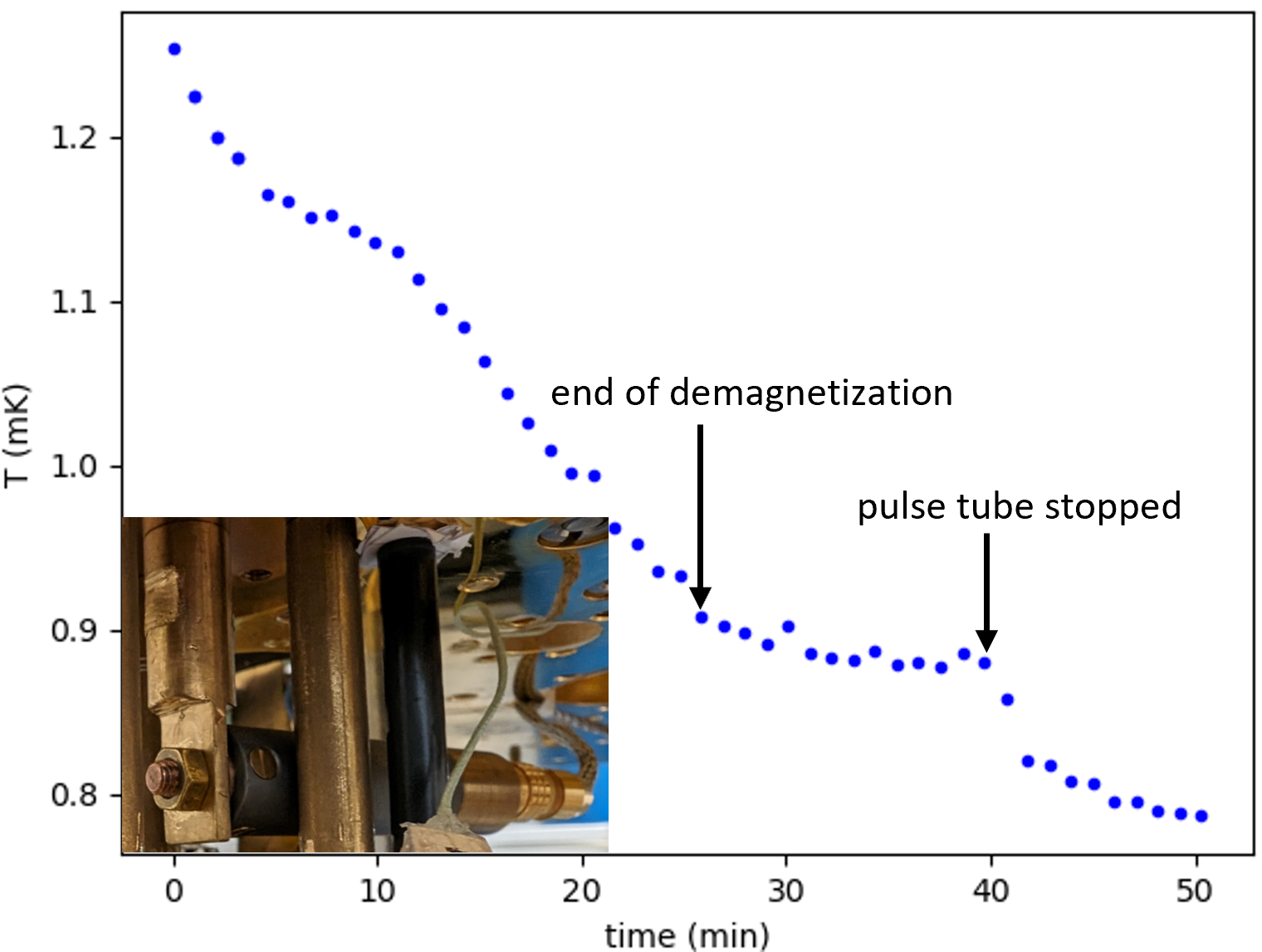}
\caption{\label{fig:mintemp} Electronic nuclear stage temperatures below 1 mK were measured with the noise thermometer (MFFT) less than six days after starting to cool from room temperature. The temperature record at the end of a demagnetization made three weeks later (main panel) showed that pulse tube operation was causing an offset in the MFFT temperature. The inset shows how the MFFT was screwed to a Au-plated Cu extension of the Cu block (see also Fig. \ref{fig:design}).}
\end{figure}

We measured temperatures below 1 mK using the MFFT attached to the Cu block less than six days after starting to cool from room temperature. In one run, we precooled to $T_{\mathrm{MFFT}}=7.7$ mK while applying $B_{\mathrm{main}}=2.84$ T using the main coil. We then demagnetized the nuclear stage to $B_{\mathrm{main}}=140$ mT (Fig. \ref{fig:mintemp}). Ten minutes after reaching the final magnetic field, the temperature stabilized at $T_{\mathrm{MFFT}}=880$ $\mu$K. We then turned off the pulse tube and observed further cooling. Ten minutes later, just before it became necessary to turn the pulse tube back on, $T_{\mathrm{MFFT}}$ had reached 790 $\mu$K and was continuing to decrease. This test was carried out at the end of the ``cooldown'', four weeks after starting to cool from room temperature. While we expect some improvement in performance with time due to the decay of the time-dependent heat leak (see below), the delay in carrying out this test involving the pulse-tube shutdown was primarily due to other practical issues. In the absence of losses, we would expect a nuclear temperature $T_{\mathrm{ideal}}=385$ $\mu$K due to the 20-fold reduction in $B_{\mathrm{main}}$. We believe that the sensitivity of $T_{\mathrm{MFFT}}$ to pulse tube vibrations could be decreased by decreasing the contact resistance between the MFFT copper body and the copper block to which it is screwed. The MFFT body is not gold plated and its tightening torque is limited by the strength of the highly pure copper from which it is constructed. The small temperature gradient along the nuclear stage and the temperature difference between the electrons and nuclei of the nuclear stage are estimated below. Furthermore, losses in the demagnetization process elevate the minimum nuclear temperature above $T_{\mathrm{ideal}}$. The losses due to vibrations and eddy current heating are quantified below.

\begin{figure}
\includegraphics[width=\linewidth]{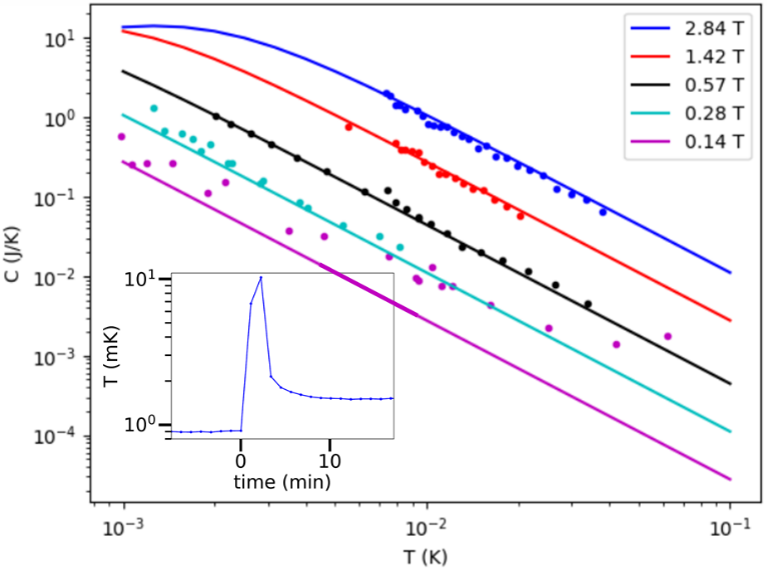}
\caption{\label{fig:C} The heat capacity of the nuclear stage (points) at different main coil currents was determined from the increase in equilibrium temperature following applied heat pulses. The measured heat capacity at high magnetic fields agreed with the theoretical nuclear heat capacity (curves) of 2 moles of aluminum at the specified fields, which were the maximum magnetic fields in the main coil corresponding to the currents. Inset: Response of the nuclear stage to a heat pulse demonstrating the short equilibration time (see main text).}
\end{figure}

We used heat pulse measurements like the one shown in Fig. \ref{fig:C} to quantify the heat capacity of the nuclear stage and place an upper limit on its thermal time constant. The pulses were applied with a heater mounted on the top weld of the Al wire bundle and the temperature was measured with the MFFT screwed to the Cu block at the other end of the wire bundle. In the measurement shown in the inset of Fig. \ref{fig:C}, $B_{\mathrm{main}}=114$ mT and 3 $\mu$W was applied to the heater for 65 seconds, causing the equilibrium value of $T_{\mathrm{MFFT}}$ to increase from 0.91 to 1.50 mK with a large transient peak. Since the heater and thermometer are located at opposite ends of the nuclear stage, the transient heating of $T_{\mathrm{MFFT}}$ to 10 mK must be due to decoupling between the electrons and nuclei of the nuclear stage and not a gradient in the electron temperature along the nuclear stage. The noise thermometer requires five minutes to reach equilibrium after the heat pulse. This upper limit on the thermal time constant of the nuclear stage is much faster than than the $\approx 100$ minute cycling time of the projected CNDR (see below).

The heat capacity measurements shown in the main panel of Fig. \ref{fig:C} were obtained from pulses like those in the inset by dividing the deposited energy by the increase in equilibrium temperature. The size of the temperature step was typically 10\% of the average temperature. The curves correspond to the theoretical nuclear heat capacity \cite{Reif65} of 2.0 mol of aluminum in the magnetic fields specified in the legend, which correspond to the applied field at the center of the main coil. The theoretical curves are in good agreement with measurements at the highest fields, so that 2.0 mol can be considered the effective quantity of aluminum in the nuclear stage. This value is close to the 2.2 mol of aluminum inside the 10 cm long main coil, and the 10\% difference may be explained by the inhomogeneity of the magnetic field. At low values of $B/T$ we observed an excess heat capacity that cannot be explained by the electronic contribution in these 2 mol of aluminum. The copper in the thermal link between the heat switch and the nuclear stage also contributes to the measured heat capacity and may be responsible for the observed excess. An anomalous heat capacity in Cu was also observed in \cite{Gloos88} and attributed to electric field gradients in the Cu due to lattice defects. While the excess heat capacity may temporarily load the proposed CNDR, we nonetheless expect the CNDR performance discussed in the conclusion since we achieved $T_{\mathrm{MFFT}}<1$ mK less than six days after cooling from room temperature.

\begin{figure}
\includegraphics[width=\linewidth]{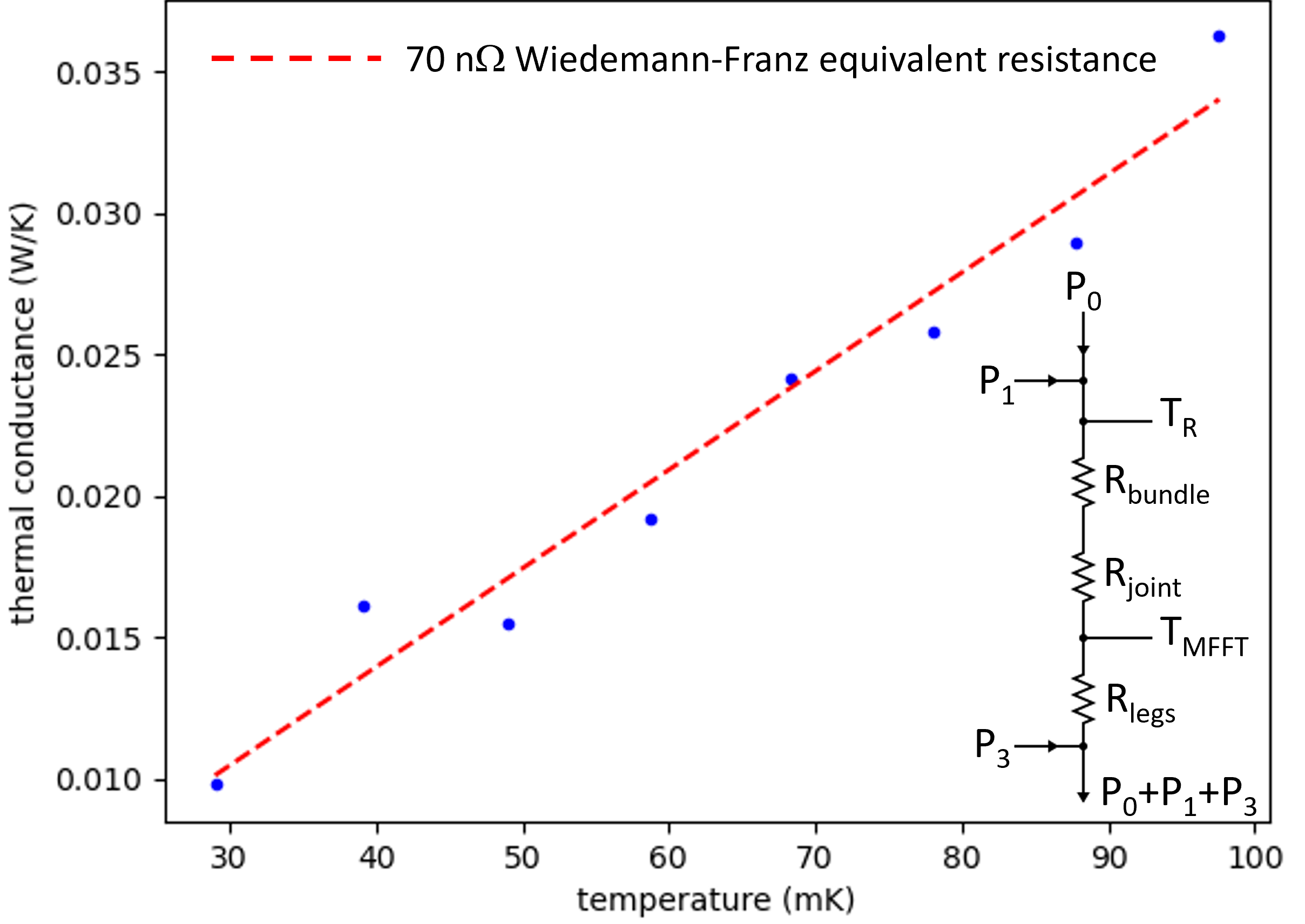}
\caption{\label{fig:kappa}Thermal conductance of the nuclear stage as a function of its temperature. Inset: Circuit depicting the thermal resistance measurement technique. $R_{\mathrm{bundle}}$, $R_{\mathrm{joint}}$ and $R_{\mathrm{legs}}$ represent thermal resistances of the Al wire bundle, Al/Au/Cu joint and Cu legs, respectively (Fig. \ref{fig:design}). For each data point in the main panel, the reading of the RuOx thermometer at the top of the nuclear stage was set to a particular value $T_{\mathrm R}$. This was achieved by applying power $P_1>0$ to the heater at the top of the stage and applying no power, $P_3=0$, to the heater at the bottom of the stage (case A), or by applying $P_1=0$ and $P_3>0$ (case B). The power $P_0$ represents the parasitic heat load. We define $R=R_{\mathrm{bundle}}+R_{\mathrm{joint}}$ and $T_{\mathrm{MFFT}}$ the temperature of the noise thermometer on the Cu block. Therefore $T_R-T_{\mathrm{MFFT}}^B=R(P_0+P_1)$ and $T_R-T_{\mathrm{MFFT}}^A=RP_0$ for cases A and B, respectively. Subtracting these equations we obtain the thermal conductance $P_1/(T_{\mathrm{MFFT}}^A-T_{\mathrm{MFFT}}^B)$ plotted in the main panel.}
\end{figure}

We also made steady state measurements of the heat transport along the nuclear stage to determine its thermal conductance. For these measurements, $B_{\mathrm{main}}=142$ mT and the small coil was energized so that all of the Al in the nuclear stage was in its normal state. The heat switch was closed. As shown in the inset of Fig. \ref{fig:kappa}, the two heater method \cite{Dhuley19} was used to determine the thermal resistance of the nuclear stage, including the contribution of the Al/Au/Cu joint. The temperature dependence of the corresponding thermal conductance is plotted in Fig. \ref{fig:kappa} along with a line showing the thermal conductance of a 70 n$\Omega$ resistance according to the Wiedemann-Franz law \cite{Ashcroft76}. This Wiedemann-Franz equivalent resistance is in good agreement with the 56 n$\Omega$ residual resistance of the nuclear stage measured electrically in liquid helium at 4.2 kelvin. Only 11 n$\Omega$ of the residual resistance was due to the Al/Au/Cu joint. For our low magnetic field heat load of $\approx 10$ nW (see below), the 70 n$\Omega$ resistance corresponds to a maximum temperature difference of 60 $\mu$K between the ends of the nuclear stage at an average temperature of 500 $\mu$K.

\begin{figure}
\includegraphics[width=\linewidth]{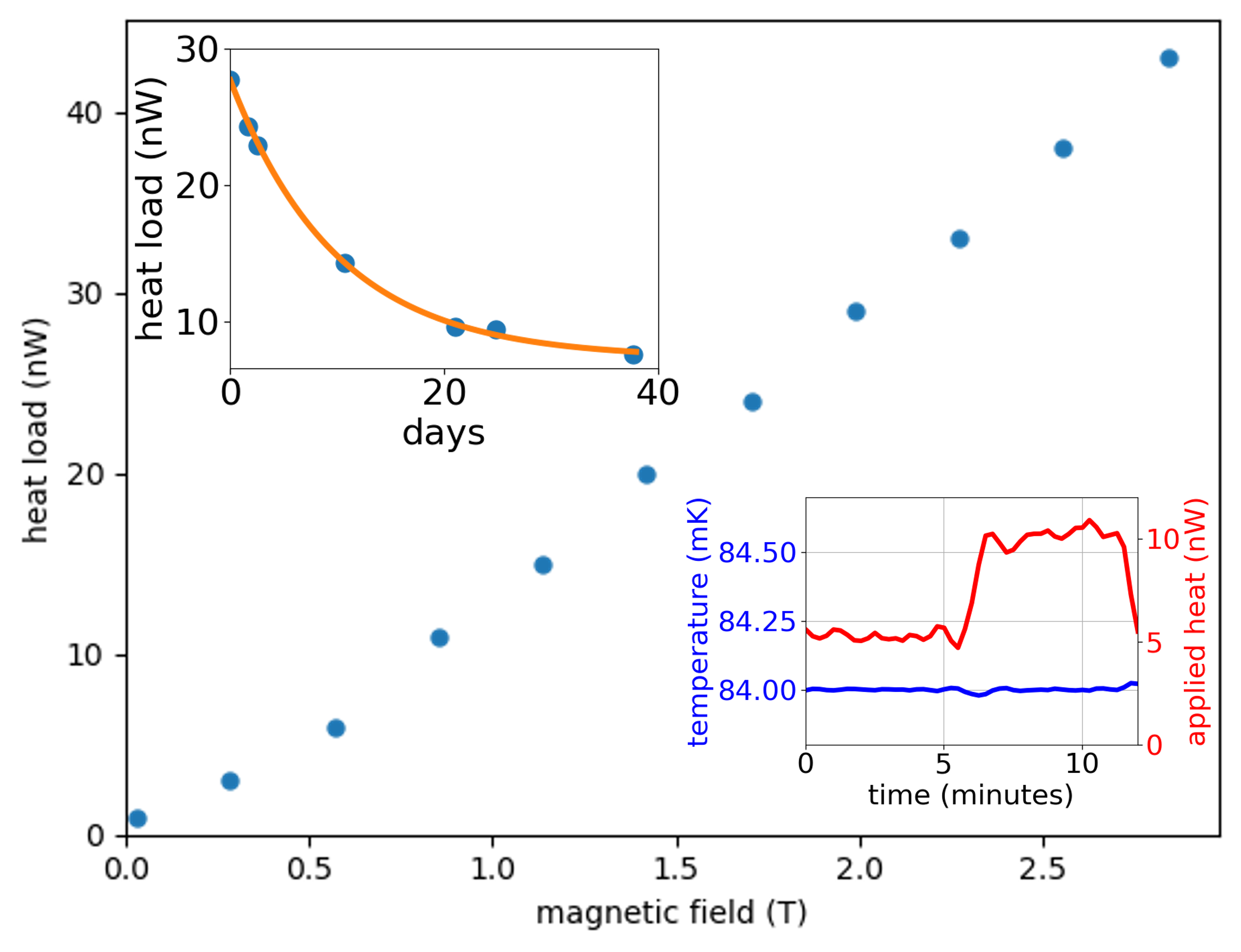}
\caption{\label{fig:static}Main: Dependence of the nuclear stage heat load on the applied magnetic field. Upper inset: The low field $B_{\mathrm{main}}=28$ mT heat load exponentially decays to a residual value of 7.3 nW with a time constant of ten days. Lower inset: The temperature of the nuclear stage was regulated at a constant value while the pulse tube was switched off and back on. The change in applied heat demonstrates that pulse tube vibrations are responsible for 5 nW of heat at low applied magnetic field.}
\end{figure}

To evaluate the parasitic heat leaks, we then measured the heat load on our nuclear stage at different constant values of $B_{\mathrm{main}}$. In order to obtain the data shown in the main panel of Fig. \ref{fig:static}, we opened the heat switch and warmed the nuclear stage to 115 mK at $B_{\mathrm{main}}=0$ using the heater at the bottom of the nuclear stage. The decrease in the heat required to maintain this temperature at different values of $B_{\mathrm{main}}$ corresponds to the heat load at constant magnetic field. We measured 29 nW at $B_{\mathrm{main}}=2$ T. This is comparable to the $\approx 40$ nW at 2 T reported by \cite{Nyeki22}, where the PrNi$_5$ is supported by a Cu structure that likely contributes to the static field heat load.

In order to determine the residual heat load in the low field limit, we opened the heat switch and used it as a known thermal resistance. In particular, we first applied heater powers to the nuclear stage that were much greater than the estimated parasitic heat load. This allowed us to verify that the open heat switch had the same temperature dependent thermal resistance as reported in \cite{Butterworth22}.  We then turned off the applied heat and allowed the system to stabilize. The heat load corresponding to the equilibrium temperature of the nuclear stage was obtained by consulting the measured heat flow as a function of the temperature of the hot side of the switch reported in \cite{Butterworth22}. The temperature of the cold side of the switch was close to the base temperature of the dilution refrigerator in both cases and was therefore negligible. The upper inset of Fig. \ref{fig:static} shows the time dependence of the heat load at $B_{\mathrm{main}}=28$ mT. Time-dependent heat leaks of a similar initial magnitude decaying approximately exponentially with a time constant of one week were previously observed in Cu and PrNi$_5$ based NDR \cite{Schwark83,Parpia85}. A portion of our time dependent heat load may come from the aluminum, which is known to release comparable amounts of heat per unit mass even in samples that are much more pure than the aluminum used here. This heat release was explained in terms of relaxation of tunneling systems \cite{Nittke96}. Furthermore, molecular hydrogen may become trapped in Cu and release heat when undergoing an ortho-para conversation at cryogenic temperatures. The time-dependent heat leak of our NDR may therefore be improved by annealing Cu parts that were not heat treated \cite{Pobell07}.

We found that a significant part of the residual heat load in the low field limit was due to pulse tube vibrations. With the heat switch open and $B_{\mathrm{main}}=28$ mT we heated the nuclear stage to 84 mK. After turning off the pulse tube we had to increase the applied heat by 5 nW to maintain the nuclear stage at this temperature (Fig. \ref{fig:static}), signifying that most of the 7.3 nW heat load that remained weeks after cooldown from room temperature was due to pulse tube vibrations. One could reduce this heat load by better isolating the pulse tube from the rest of the cryostat, but we expect high performance of an aluminum CNDR even without doing so (see below).

\begin{figure}
\includegraphics[width=\linewidth]{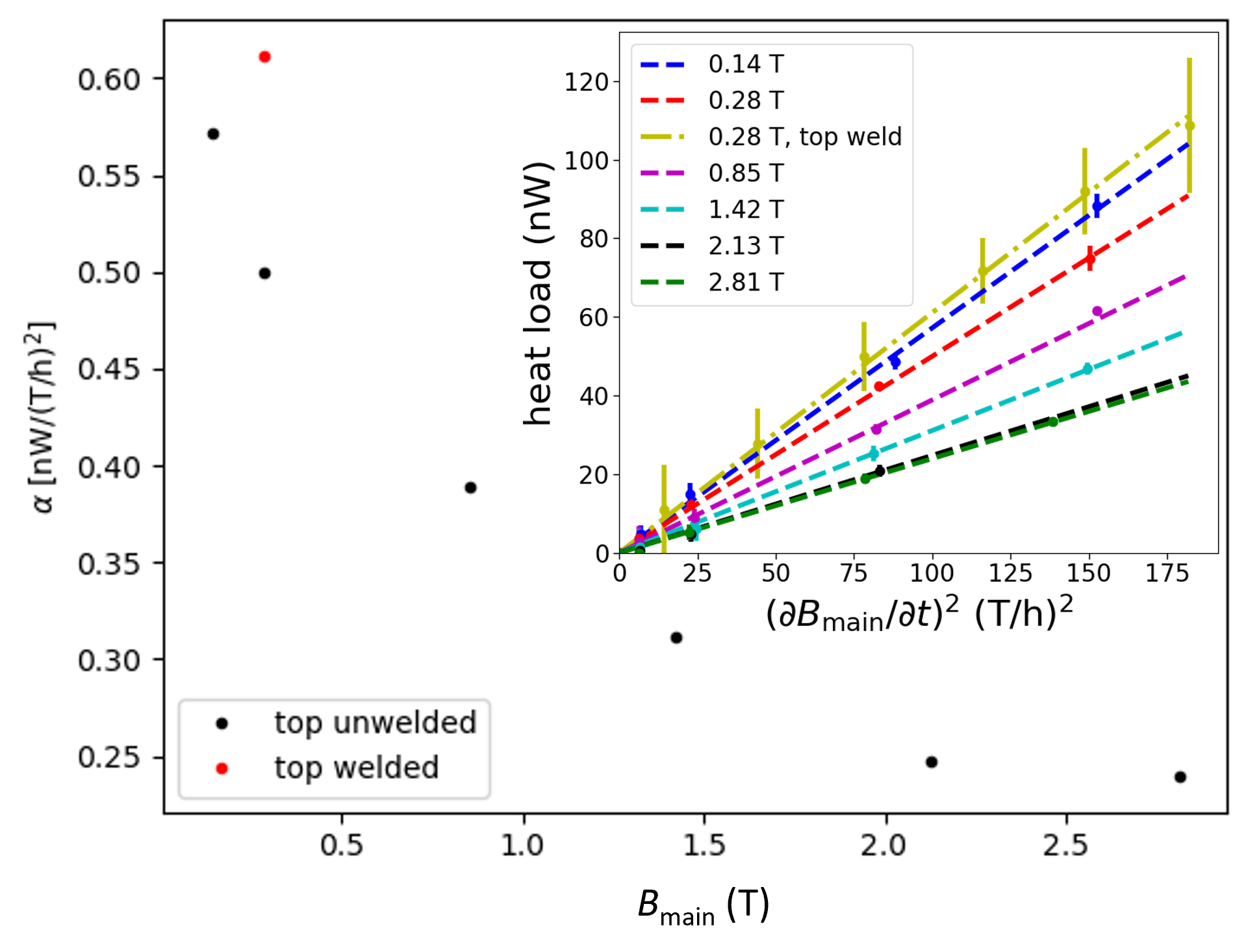}
\caption{\label{fig:dBdt}Inset: Heat load due to $B_{\mathrm{main}}$ oscillating around the average $B_{\mathrm{main}}$ shown in the legend. All measurements were made before the addition of the top weld of the Al wire bundle (see main text) except the one labelled ``top weld''. The heating is higher in the latter case due to the added aluminum. Vertical lines are error bars and dashed lines are fits of the form $\alpha(\partial B/\partial t)^2$. The coefficient $\alpha$ is plotted as a function of the average $B_{\mathrm{main}}$ in the main panel.}
\end{figure}

The contribution to eddy current heating that would be present in the absence of vibrations was quantified by applying an oscillating magnetic field to the nuclear stage. In particular, the stage temperature was regulated at 130 mK with the heat switch open at a particular static field $B_{\mathrm{main}}$. We then oscillated $B_{\mathrm{main}}$ with amplitude 28 mT around the initial static field. The resulting decrease in the heater power required to maintain the stage at 130 mK corresponded to the eddy current heating. These results are plotted in the inset of Fig. \ref{fig:dBdt}. As expected, we observed a quadratic dependence of eddy current heating on $\partial B/\partial t$. The corresponding heating coefficient is plotted as a function of average $B_{\mathrm{main}}$ in the main panel of Fig. \ref{fig:dBdt}. We believe that this coefficient decreases with average $B_{\mathrm{main}}$ due to magnetoresistance of the aluminum, which, to our knowledge, has not been measured in the low mK range \cite{Fickett71}. While magnetoresistance of copper has a large influence on eddy current heating in systems such as the one described in \cite{Nyeki22}, where electrical resistance increases by a factor of over 20 in 6 tesla, the magnetoresistance of aluminum is different \cite{Fickett72}. The maximum demagnetization rate in the present work was 5.1 T/h, so that this contribution to the heat load was always below $\approx 20$ nW. In another cooldown, we characterized the non-adiabaticity of the NDR by determining the equilibrium value of $B_{\mathrm{main}}/T_{\mathrm{MFFT}}$ in a temperature range where decoupling of the thermometer is negligible. In particular, we started at $T_{\mathrm{MFFT}}=6.73$ mK, yielding $B_{\mathrm{main}}/T_{\mathrm{MFFT}}=0.42$ T/mK at the initial $B_{\mathrm{main}}=2.84$ T. We demagnetized at 1.6 T/h, pausing for 1 hour after each 0.28 T decrease in $B_{\mathrm{main}}$. When $T_{\mathrm{MFFT}}$ reached 3.0 mK, $B_{\mathrm{main}}/T_{\mathrm{MFFT}}$ had decreased to 0.38 T/mK.

\section{Conclusion}
\begin{figure*}
\includegraphics[width=\linewidth]{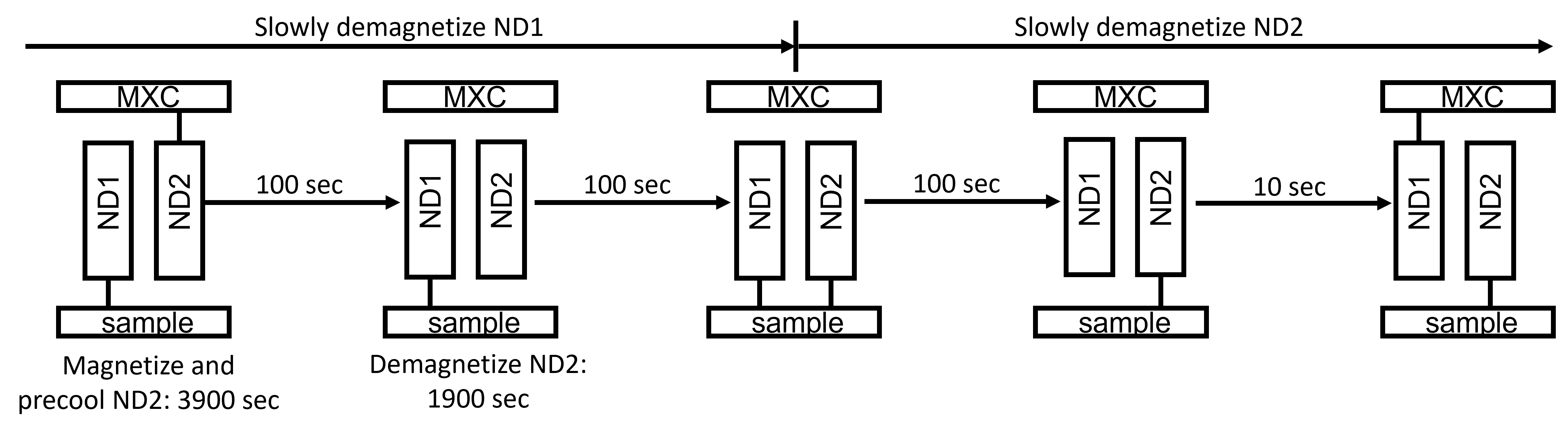}
\caption{\label{fig:parallel} Half a cycle of the parallel CNDR configuration, indicating the maximum time for each step. ND1(2) stands for nuclear demagnetization stage 1(2). Closed heat switches linking the nuclear stages to the mixing chamber (MXC) or sample plate are indicated by vertical lines. The times shown over arrows in between states of the CNDR correspond to the delay for toggling a heat switch, which is limited by eddy current heating of the switch. The effect of such heating is less significant when connecting a ND to the MXC just before regeneration, so the switching delay is much shorter.}
\end{figure*}

We have characterized the performance of a new kind of nuclear demagnetization stage with a short thermal time constant and low susceptibility to parasitic heating. We conclude with an estimate of the cooling power of a CNDR based on two copies of the individual nuclear stage demonstrated here. We consider the parallel CNDR configuration \cite{Schmoranzer20} (Fig. \ref{fig:parallel}), so that the cooling power is the amount of heat that can be absorbed after connecting a freshly cooled stage to the sample divided by the regeneration time of the other stage.

Estimating the absorption capacity depends on an accurate measurement of the nuclear temperature. We verified that our thermometer was sufficiently well coupled so that $T_{\mathrm{MFFT}}$ differed from the nuclear temperature by only a few percent at $T_{\mathrm{MFFT}}=2.4$ mK. We could then use magnetic field ratios to place limits on the minimum nuclear temperature. In particular, when $B_{\mathrm{main}}$ was changed from 568 mT to 142 mT and then back to 568 mT, the temperatures at the start and end of this sequence were respectively 2.03 and 2.44 mK. This implies that the nuclear temperature at $B_{\mathrm{main}}=142$ mT was between 510 and 610 $\mu$K, so that we can estimate $B_{\mathrm{main}}/T=0.254$ T/mK. With 2.0 mol of Al, this corresponds to a removed entropy of 0.438 J/K \cite{Reif65}. The maximum heat that can be absorbed by the nuclear system at 560 $\mu$K is then 245 $\mu$J.

Quickly achieving $T<1$ mK, starting from room temperature, may be important in some applications. In a subsequent cooldown of the same NDR, after adjusting the centering screws (Fig. \ref{fig:design}), we achieved somewhat higher absorption capacity less than six days after starting to cool from room temperature, measuring $T_{\mathrm{MFFT}}=1.85$ mK at $B_{\mathrm{main}}=568$ mT.

The regeneration time of one of our stages has several contributions (Fig. \ref{fig:parallel}). To achieve $B_{\mathrm{main}}=142$ mT and $T=560$ $\mu$K, the upper limits are as follows: 300 seconds for toggling the four heat switches and 1900 seconds each for magnetization and demagnetization of the nuclear stage. Based on the measured cooling power of our dilution refrigerator, we estimate 2000 seconds for pre-cooling to 7 mK at $B_{\mathrm{main}}=2.84$ T, yielding a total regeneration time of 6100 seconds. Our estimate relies on the fact that the effective electrical resistance between the nuclei and the mixing chamber during the pre-cool will be only a few tens of n$\Omega$, so that the delay is limited by the cooling power of the dilution refrigerator \cite{Blondelle14}.

We therefore expect a cooling power of at least 40 nW at 560 $\mu$K less than six days after cooling from room temperature. The low field heat leak measured in the present work, which decreased from 30 to 7 nW, would yield a small decoupling of the electrons from the nuclei ranging from 115 to 27 $\mu$K \cite{Lounasmaa74}. We therefore anticipate that a CNDR based on Al stages of the type described here will maintain temperatures well below 1 mK indefinitely.

\begin{acknowledgments}
We acknowledge support from the ERC StG grant UNIGLASS (Grant No. 714692) and ERC PoC grant NewCooler (Grant No.
101113096). The research leading to these results received funding from the European Union’s Horizon 2020 Research and Innovation Program under Grant Agreement No. 824109. Deposition of the Au film took place at the Plateforme Technologique Amont (PTA) of Grenoble. The work was partly supported by the French Renatech network.
\end{acknowledgments}

\end{document}